\documentclass[aps]{revtex4}

\usepackage{graphicx}
\newcommand{\be}{\begin{equation}}
\newcommand{\ee}{\end{equation}}
\newcommand{\ben}{\begin{eqnarray}}
\newcommand{\een}{\end{eqnarray}}
\newcommand{\lb}{\label}

\begin{document}

\title{Finite-size effects on the phase structure of the Nambu-Jona-Lasinio model}

\author{L. M. Abreu}
\email{lmabreu@fma.if.usp.br}
\author{M. Gomes}
\email{mgomes@fma.if.usp.br}
\author{A. J. da Silva }
\email{ajsilva@fma.if.usp.br}

\affiliation{\it Instituto de F\'{i}sica, Universidade de S\~{a}o Paulo, Caixa Postal 66318, 
CEP 05315-970, S\~{a}o Paulo-SP, Brazil}

\begin{abstract}
The Nambu-Jona-Lasinio (NJL) model is one of
the most frequently used four-fermion models in the study of dynamical symmetry breaking. 
In particular, the NJL model is convenient for that analysis at finite
temperature, chemical potential and size effects, as has
been explored in the last decade. 
With this motivation, we investigate the finite-size effects on the phase structure of
the NJL model in $D = 3$ Euclidean dimensions, in the situations that one, two and three
dimensions are compactified. In this
context, we employ the zeta-function and
compactification methods to calculate the effective potential and gap
equation. The critical lines that separate trivial and non-trivial
fermion mass phases in a second order transition are obtained. 
We also analyze the system at finite temperature, considering the
inverse of temperature as the size of one of
the compactified dimensions.  \\

\noindent PACS number(s): 11.10.Kk; 11.10.Wx; 11.30.Rd\\

\end{abstract}
\maketitle

\section{Introduction}

The last decades witnessed  significant investigations on the phase
structure of quantum field theories, in particular on the chiral
symmetry phase transitions in Quantum Cromodynamics (QCD). However,
due to its complex structure, effective models have been largely
employed to simplify that analysis. Among them, one of the most frequently used
is the four-fermion theory  known as Nambu-Jona-Lasinio (NJL) model
\cite{NJL}. The NJL model is specially convenient for the
investigation of dynamical symmetries when the system is under certain
conditions, like finite temperature, finite chemical potential, external gauge field, gravitation field and others \cite{Kl,HK,Bu}. 

Finite-size effects on the phase transitions of four-fermion 
models have also attracted a great interest \cite{Kim,KS,KKK}.
This question emerges when the system has a
finite size and  it is not clear if it is large enough to apply the
thermodynamic limit in a usual way; frequently it is necessary to take into account the
fluctuations due  to finite-size effects.
In particular, Ref.~\cite{KS} performed a numerical investigation of a three-dimensional
four-fermion model in a finite-size scaling analysis,
where the finite-size effects act as an external field. On the other
hand, Ref. \cite{KKK} studied the NJL model in the framework of the
multiple reflection expansion, where in terms of a modified density of states finite-size effects are
included (see also Ref. \cite{HCGL}). The critical
temperature was suggested to decrease as the system is reduced. 


In this paper, we investigate finite-size effects on the dynamical
symmetry breaking in a different way. We study the Euclidean
three-dimensional NJL model in the framework of zeta-function and
compactification methods \cite{EE}. This procedure in principle allows
us to explore the mentioned model with one, two or three compactified
dimensions with antiperiodic boundary conditions \footnote{For a discussion of periodicity of boundary conditions see \cite{Kim}.} and compare their effects in the phase diagram of the
model. With the choice of all dimensions being spatial, the system is considered confined between two parallel
planes a distance $L$ apart, confined to a infinity cylinder having a square transversal section 
of area $L^2$, and to a cubic box of volume $L^3$, for one, two and three compactified coordinates, respectively. At
finite temperature, we associate one of the 
compactified coordinates to the range $[0, \beta]$, where $\beta$ is
the inverse of the temperature $T$. 
In this setting, we calculate and determine analytically 
the size-dependence of the effective potential and gap equation. 
Phase diagrams, where the symmetric and broken phases are separated by
a size-dependent critical line, are obtained. 

This paper is organized as follows. In Section II, we briefly review
the NJL model and obtain the effective potential in mean-field
approximation. In Section III, we apply the zeta-function
method to derive the effective potential. After,
Section IV is devoted to analyze the size-dependent gap equation. 
The phase diagrams are shown and discussed in Section~V.
Finally, in Section VI are given a summary and concluding remarks.

\section{The formalism}

\subsection{ NJL model as an effective theory of QCD}


QCD is the theory of strongly interacting matter, which is constituted
of quarks and gluons. In the study of dynamical symmetry breaking, the
attractive interaction between quarks and antiquarks, coming from
complicated processes of exchanges of gluons, are often replaced by an
effective interaction between them. This interaction generates the
quark-antiquark condensation in  the vacuum when the interaction
exceeds a critical strength, and this condensation is responsible for
the dynamical symmetry breaking. Thus, theories incorporating this
mentioned symmetry, such as the NJL model, are very useful in this analysis. 

The NJL model can be obtained heuristically as follows. Inspired by
the QCD, let us consider the following approximation for the action  which describes the interactions between quarks mediated by exchange of one gluon, 
\ben 
{\cal A }_{exchange} = \int d^4 x \left[ \bar{q} \; \left(i \not{\!\partial} -M + g A_{a}^{\mu} \lambda_{c}^{a} \gamma _{\mu} \right) \; q - \frac{1}{2} \int d^4 y \; A_{a}^{\mu} (x)\left[ D_{ab}^{\mu \nu} (x-y)\right]^{-1} A_{b}^{\nu} (y) \right] ,
\lb{GCM}
\een
where the spinors $q$ and $\bar{q}$ represent the quark and antiquark
fields carrying $N_f$ flavors and $N_c$ colors, $A_{a}^{\mu}$ is the
gluon field, $\lambda_{c}^{a}\;(a=1...(N_c^2 -1))$ are the generators
of ${\rm{SU}} (N_c)$ and $M$ is the quark mass matrix. The quantity $D_{ab}^{\mu \nu} (x-y)$ is interpreted as the effective gluon propagator. Note that there is no self-gluon interaction term. 
The functional integration over the gluon field in Eq. (\ref{GCM}) leads to 
\ben 
\bar{ {\cal A}} _{exchange} = \int d^4 x \left[ \bar{q} \; \left(i \not{\!\partial} -m  \right) \; q
 + \frac{g^2}{2} \int d^4 y \; \left( \bar{q}\lambda_{c}^{a} \gamma _{\mu} q \right) (x) D_{ab}^{\mu \nu} (x-y) \left( \bar{q}\lambda_{c}^{b} \gamma _{\nu} q \right) (y)\right] .
\lb{GCM2}
\een

Next, we assume that the interaction obeys to the Feynman gauge, and
moreover it is a contact interaction, which means 
$ g^2 D_{ab}^{\mu \nu} (x-y) \propto G \delta _{ab}g^{\mu \nu } \delta
(x-y)$.  Furthermore, we can use the Fierz identities for color,
flavor and  Dirac spaces \cite{Kl,BB}, to rewrite the interaction part
of the Lagrangian in the action explicited in Eq. (\ref{GCM2})  as (considering only the $\bar{q} q$-current), 
\ben 
{\cal L}_{int} \propto G   \left( \bar{q} \Lambda^{\alpha} q \right)\left( \bar{q} \Lambda_{\alpha} q \right) , 
\lb{GCM3}
\een
where $\Lambda^{\alpha} = I_c \otimes \lambda_f ^i \otimes K^j$, with $\lambda_f ^i= \left( I/\sqrt{N_f}, \lambda_f ^a \right); \; a=1,..., N_f^2 -1$ being the generators of $U(N_f)$, and $K^j = (I, i\gamma _5, \frac{i\gamma ^{\mu}}{\sqrt{2}}, \frac{i\gamma ^{\mu}\gamma _5}{\sqrt{2}})$.

Eq. (\ref{GCM3}) characterizes the interaction part of the Lagrangian
of the NJL model. It simplifies a gluon exchange interaction between quarks and
antiquarks to a local interaction, i.e. a four-quark point interaction. 

\subsection{ The effective potential}

In the previous subsection we have heuristically derived the NJL model
from a simplified version of QCD. To analyze With its dynamical
symmetry phase diagram, we will begin by deriving its effective
potential. For generality, we will consider the NJL model as an
independent model by itself: the coupling constant is considered to be
and arbitrary parameter, without any a priori relation with QCD aspects. Also, the dimension of the system is considered here arbitrary ($2\leq D<4$), as well as we limit ourselves to mesonic scalar and pseudoscalar modes. 

We will treat with the colorless and massless NJL model, 
\be
{\cal L} = \bar{q} \; i \not{\!\partial} \; q + \frac{G}{2} \sum ^{N^2-1}_{i=0} \left[ \left( \bar{q} \lambda ^i q \right)^2 + \left( \bar{q} i \gamma _5 \lambda ^i q \right)^2 \right],
\lb{NJL}
\ee
where the spinors $q$ and $\bar{q}$ here carry $N$ flavors, and the matrices $\lambda ^i$ are the generators of the $U(N)$, $\lambda ^i= \left( I/\sqrt{N}, \lambda ^a \right); \; a=1,..., N^2 -1$. 

In order to study the phase structure of this model, it is convenient to perform the bosonization. We introduce auxiliary fields $M^i$ and $\phi ^i$ through the following way, 
\ben
1 & = & \int {\cal D}M^i \; \delta{(M^i - \bar{q} \Gamma ^i q)} \nonumber \\
& = &  \int {\cal D}M^i {\cal D}\phi ^i \exp{\left[ i \int d^D x (M^i - \bar{q} \Gamma ^i q) \phi ^i \right] }, 
\lb{unity}
\een
where $\phi ^i = (\sigma^i, \pi^i) $ and $\Gamma ^i =(\lambda ^i, i \gamma _5 \lambda ^i)$. In this sense, $\sigma^i $ and $\pi^i $ are fields corresponding to the scalar and pseudoscalar bilinear forms, respectively. The use of auxiliary fields in the generating functional of the model introduced in Eq. (\ref{NJL}) yields 
\be
{\cal Z} \propto \int {\cal D}\bar{q} {\cal D}q {\cal D}\phi ^i
\exp \left\{ i \int d^D x \left[  \bar{q}\left( \; i \not{\!\partial} \; - \Gamma ^i \phi ^i \right) q - \frac{\left( \phi ^i \right)^2}{2G}  \right] \right\}. 
\lb{gen1}
 \ee
where the source terms have been omitted. Thus, the integration over the Grassmann variables $q$ and $\bar{q}$ results in 
\be
{\cal Z} \propto \int {\cal D}\phi ^i
\exp \left\{ i {\cal A}_{eff} \right\}, 
\lb{gen2}
 \ee
where
\be
{\cal A}_{eff}= -\int d^D x \frac{\left( \phi ^i \right)^2}{2G} - i \; \rm{Tr} \ln{\left( \;\frac{ \; i \not{\!\partial} \; - \Gamma ^i \phi ^i}{\mu} \right)} 
\lb{eff}
\ee
is the effective action associated to the model described in
Eq. (\ref{NJL}) with one-loop corrections; $\mu $ is a constant with
dimension of mass. Here $\rm{Tr}$ means the trace over coordinate,
Dirac and internal spaces. 

In the above context we restrict $\phi$ to be independent of the
spatial coordinates; moreover, we impose $\pi^i =0$, $\sigma ^a
=0\;(a=1,..., N^2 -1)$; only the scalar part of $\phi$ which is
diagonal in flavor space, $ \sigma ^0$, can assume non-vanishing
values. Thus, it is convenient to define $\lambda ^0 \sigma ^0 \equiv
\sigma$, with $\sigma$ assuming the role of dynamical fermion mass, 
its non-trivial value meaning that the system would be in the broken phase. 

Taking into account the above considerations and performing the Wick
rotation in the $x^0$-coordinate, we can use Eq. (\ref{eff}) to obtain the the effective potential, 
\be
\frac{1}{N} U_{eff}  =  -\frac{{\cal A}_{eff}}{V}
 =  \frac{ \sigma ^2}{2G} + U_{1} (\sigma), 
\lb{effpot}
\ee
where $V$ is the volume, and
\be
U_{1} (\sigma) = - h  \int \frac{d^D p}{(2\pi)^D} \ln{\left( \; \frac{k_E^2 + \sigma^2}{\mu ^2}  \right)},
\lb{oneloop}
\ee
with $h$ being the dimension of the Dirac representation. 

As our interest is to describe the finite-size effects on the phase structure of this model, it is convenient to use the zeta-function regularization techniques, described below.

\section{The zeta-function regularization approach}

\subsection{The zeta-function}

This method is based on the fact that one can define a generalized zeta-function from the eigenvalues $\alpha_i$ of a differential operator $A$ of order 2 on a $d$-dimensional compact manifold (a bounded manifold), 
\be 
\zeta _A (s) = \sum _i \alpha _i ^{-s}, 
\lb{zetaA}
\ee
valid for $\rm{Re} \; s > \frac{d}{2}$. For the other values of $s$, we must perform an analytical continuation (for a review, see \cite{EE,KK}). 

In particular, 
\be 
\zeta _A' (0) =\left. \frac{d \zeta _A (s)}{d s} \right|_{s=0} = - \sum _i \ln{\alpha _i} = -\rm{Tr} \ln{A}. 
\lb{zeta0}
\ee 
Therefore one can associate the operator $A $ to the one appearing in the argument of the logarithm in Eq. (\ref{oneloop}), allowing to rewrite $U_{1} (\sigma)$ as 
\be
U_{1} (\sigma) =  \frac{M}{2V}  \left[ \zeta _A' (0) + \ln{\mu ^2}\zeta _A (0)\right].
\lb{U1}
\ee

\subsection{Compactification}

At this point we will take into account the finite size effects. Considering the system restricted to have $d \leq D$ compactified dimensions, we can rewrite the coordinate vectors as $x_E = (y,z)$, where 
\be
y=\left( y_1 \equiv x_E ^1, ..., y_n \equiv x_E ^n  \right), \;\; z=\left( z_{1} \equiv x_E ^{n+1}, ..., z_{d} \equiv x_E ^{D}  \right),
\ee
with the $z_i$ components defined in the domain $z_i \in [0, L_i]$ and $n=D-d$. The compactification of the coordinates makes the $k_z$-components of momenta to assume discrete values, 
\ben
 k_{z}^i \rightarrow \frac{2\pi}{L_i}(n_{i} + c_i ), \; n_i = 0, 1, 2, \ldots ; 
\lb{prescription}
\een
where $c_i = \frac{1}{2} \; (i=1,2,...,d)$ for antiperiodic boundary conditions. Notice that we may associate a given $L_i $  to the inverse of temperature $\beta = 1/T$, i.e. $L_i \equiv \beta$; in this situation the chemical potential $\mu_0$ also could be introduced through the rule $c_i=\frac{1}{2} - \frac{i\beta \mu_0}{2 \pi}$.

The prescription given in Eq. (\ref{prescription}) makes the zeta-function in (\ref{zeta0}) be given by 
\be
\zeta _A (s) =  V_n
\sum_{n_{1},...,n_{d} =  -\infty}^{+\infty }
\int \frac{d^n k_y}{(2\pi)^{n}}\left[k_z ^2 + k_y ^2+\sigma^{2} \right]^{-s },
\lb{zeta1}
\ee
where $V_n$ is the $n$-dimensional volume. The techniques of dimensional regularization can be used to perform the integration over the $k_y$-components, yielding 
\be
\zeta _A \left( s;\left\{a_i\right\}, \left\{c_i\right\} \right) =  \frac{V_n}{(4\pi)^{n/2}} \frac{\Gamma \left( s- \frac{n}{2} \right)}{\Gamma \left( s \right) }
Y_{d}^{\sigma^2}\left( s-\frac{n}{2};\left\{a_i\right\}, \left\{c_i\right\} \right),
\lb{zeta2}
\ee
where $Y_{d}^{\sigma^2} \left( \nu ; \left\{a_i\right\}, \left\{c_i\right\} \right)$ is the generalized Epstein-zeta function, defined by 
\be
Y_{d}^{\sigma^2}\left( \nu ; \left\{a_i\right\}, \left\{c_i\right\} \right) = \sum_{n_{1},...,n_{d} = -\infty}
^{+\infty }\left[ a_{1} \left(n_{1} + c_1 \right)^{2}+\cdots +a_{d} \left(n_{d} + c_d \right)^{2}+\sigma^{2} \right]^{-\nu }.
\lb{epstein1}
\ee
being $a_i = \frac{4 \pi ^2}{L_i ^2}$; notice that $Y_{d}^{\sigma^2}$ is valid for $\rm{Re} \; \nu > \frac{d}{2}$.

\subsection{The pole structure of the zeta-function}

According to Eq. (\ref{U1}), in order to obtain the effective
potential, we must perform the derivative  of the $\zeta _A \left(
  s;\left\{a_i\right\}, \left\{c_i\right\} \right)$ at $s=0$.
However, we note from Eq. (\ref{zeta2}) that there is a different
pole structure coming from  $\frac{\Gamma \left( s- \frac{n}{2}
  \right)}{\Gamma \left( s \right) }$ at $s=0$ for $n$ even or
not. So, we are forced to consider these different
situations. Furthermore, the function $Y_{d}^{\sigma^2}$ also has a
different behavior for $d$ even or odd, as it is discussed in
Ref. \cite{KK}. In the case of $d$ odd, $Y_{d}^{\sigma^2}$ has poles
at $\nu = \frac{d}{2}, \frac{d}{2} -1,..., \frac{1}{2}$ and
$-\frac{(2l+1)}{2}$ ($l \in  \rm{\textbf{N}}_0 $), while for $d$ even
they are at $\nu = \frac{d}{2}, \frac{d}{2} -1,..., 1$. Nevertheless, as it
will be shown in the next section, for the study of the phase diagram
we take $\sigma ^2 \rightarrow 0$; in this limit, $Y_{d}^{\sigma^2}$
has pole only at $ \frac{d}{2}$. This fact assures that for further
calculations the pole structure of
$Y_{d}^{\sigma^2}$ is not relevant.

Then, from Eq. (\ref{zeta2}) we have to deal with the following situation,
\ben
\zeta _A' \left( 0;\left\{a_i\right\}, \left\{c_i\right\} \right) & = &
\frac{V_n}{(4\pi)^{n/2}} \lim_{s \rightarrow 0} \left[\frac{d}{d s}\left(\frac{\Gamma \left( s- \frac{n}{2} \right)}{\Gamma \left( s \right) }\right)
Y_{d}^{\sigma^2}\left( s-\frac{n}{2};\left\{a_i\right\}, \left\{c_i\right\} \right) \right. \nonumber \\
& + & \left. \frac{\Gamma \left( s- \frac{n}{2} \right)}{\Gamma \left( s \right) }
\frac{d}{d s} \left(Y_{d}^{\sigma^2}\left( s-\frac{n}{2};\left\{a_i\right\}, \left\{c_i\right\} \right)\right)\right], 
\lb{zetader}
\een
After the necessary manipulations, we obtain 
\ben
\zeta _A' \left( 0;\left\{a_i\right\}, \left\{c_i\right\} \right) & = &
\frac{V_n}{(4\pi)^{n/2}}\frac{(-1)^{\frac{n}{2}}}{\frac{n}{2}!} \left\{ \left.\frac{d}{d s}Y_{d}^{\sigma^2 }  \left(s -\frac{n}{2};\left\{a_i\right\}, \left\{c_i\right\} \right) \right|_{s=0} \right. \nonumber \\
&- & \left. Y_{d}^{\sigma^2}\left( -\frac{n}{2};\left\{a_i\right\}, \left\{c_i\right\} \right)\left[ \gamma + \psi \left(\frac{n}{2} + 1 \right) \right] \right\}, \;\; \rm{ for \;\textit{n} \; even,}
\lb{zetaeven}
\een
where $\gamma $ is the Euler-Mascheroni constant and $\psi \left(k
\right)$ is the digamma function, and 
\ben
\zeta _A' \left( 0;\left\{a_i\right\}, \left\{c_i\right\} \right) & = &
\frac{V_n}{(4\pi)^{n/2}} \Gamma \left( - \frac{n}{2} \right) Y_{d}^{\sigma^2}\left( -\frac{n}{2};\left\{a_i\right\}, \left\{c_i\right\} \right), \;\;\rm{ for \;\textit{n} \; odd}.
\lb{zetaodd}
\een
Due to the discussion of the pole structure above, notice that Eq. (\ref{zetaodd}) is valid for $d$ even or in
the limit $\sigma ^2 \rightarrow 0$. 

Now, coming back to Eq. (\ref{U1}), we have two possibilities for $U_1
(\sigma)$: the use of Eq. (\ref{zetaeven}) gives
\ben
U_{1} \left(\sigma ;\left\{a_i\right\}, \left\{c_i\right\} \right) & = &  \frac{h}{2V_d (4\pi)^{n/2} }
\frac{(-1)^{\frac{n}{2}}}{\frac{n}{2}!} \left\{ Y_{d}^{\sigma^2 \;'}  \left( -\frac{n}{2};\left\{a_i\right\}, \left\{c_i\right\} \right) \right. \nonumber \\
&+ & \left. Y_{d}^{\sigma^2}\left( -\frac{n}{2};\left\{a_i\right\}, \left\{c_i\right\} \right)\left[ \ln{\mu ^2} - \gamma - \psi \left(\frac{n}{2} + 1 \right) \right] \right\}, \;\;\rm{ for \;\textit{n} \; even},
\lb{U1even}
\een
while with the help of Eq. (\ref{zetaodd}) we obtain 
\be
U_{1} \left(\sigma ;\left\{a_i\right\}, \left\{c_i\right\} \right) =  \frac{h}{2V_d (4\pi)^{n/2} } \Gamma \left( - \frac{n}{2} \right) Y_{d}^{\sigma^2}\left( -\frac{n}{2};\left\{a_i\right\}, \left\{c_i\right\} \right), \;\;\rm{ for \;\textit{n} \; odd},
\lb{U1odd}
\ee
valid for $d$ even or in
the limit $\sigma ^2 \rightarrow 0$.

Now we need to perform the analytical continuation of the generalized Epstein zeta-function $Y_{d}^{\sigma^2}\left( -\frac{n}{2};\left\{a_i\right\}, \left\{c_i\right\} \right)$. Making use of Mellin transforms and relations between Jacobi theta-functions \cite{EE}, we find 
\begin{eqnarray}
Y_{d}^{\sigma^2}\left( \nu;\left\{a_i\right\}, \left\{c_i\right\} \right) & =& \frac{\pi
^{\frac{d}{2}}}{\sqrt{a_1 \ldots a_d}\Gamma (\nu )}\left[ \Gamma \left(\nu -\frac{d}{2}\right) \left( \sigma
\right)^{d-2\nu } + \sum_{\left\{n_{j}\right\}\in \rm{\textbf{Z}}}
\left( \frac{\pi}{\sigma}
\sqrt{ \sum_{j=1}^{d} \frac{n_{j}^{2}}{a_{j}}}\right)^{\nu - \frac{d}{2}}\right.  \nonumber \\ & \times& \left. \cos{(2\pi n_1 c_1)}\ldots \cos{(2\pi n_d c_d)}
K_{\nu - \frac{d}{2}} \left( 2 \pi \sigma
\sqrt{ \sum_{j=1}^{d}  \frac{n_{j}^{2}}{a_{j}}}
\right) \right] , 
\label{epstein2}
\end{eqnarray}
where $\rm{\textbf{Z}}$ means the set of non-vanishing integers.

For small values of $\sigma^2$, $\sigma^2<<1$, it
is possible to use the binomial expansion for $ Y_{d}^{\sigma^2}$ defined in Eq. (\ref{epstein1}) to expand it in powers of the $\sigma$-field , obtaining the expression
\be
Y_{d}^{\sigma^2}\left( q;\left\{a_i\right\}, \left\{c_i\right\} \right) =  \sum_{j=0}^{\infty}
\frac{(-1)^{j}}{j!} T(q,j)  Y_{d}\left( q+j; \left\{a_i\right\}, \left\{c_i\right\} \right) \sigma^{2j}, 
\label{epstein3}
\ee
where $T(q,j)= \frac{ \Gamma (q+j)}{\Gamma (q)}, 
$ and
\be
Y_{d}\left( \nu; \left\{a_i\right\}, \left\{c_i\right\} \right) = \sum_{\{ n_i\}\in \rm{\textbf{Z}}_0}
  \left[ a_{1}\left(n_{1}+ c_1 \right)^{2} +...+a_{d}\left(n_{d}+ c_d \right)^{2} \right]^{-\nu}
\label{eps}
\ee
is the homogeneous generalized Epstein zeta-function, with $\rm{\textbf{Z}}_0$ being the set of integers.

\section{The gap equation}

\subsection{Case without boundaries}

In this section we focus our attention to the gap equation, which will allow us to analyze the phase structure of the model. It is obtained by minimizing the effective potential with respect to $\sigma$, 
\be 
\left.\frac{\partial  }{\partial \sigma}U_{eff} \left(\sigma ;\left\{a_i\right\}, \left\{c_i\right\} \right)\right|_{\sigma = m} = 0, 
\lb{gap}
\ee
where $m$ is the dynamical mass of the fermion. 

For completeness, we briefly describe the case without boundaries (for reviews see Refs. \cite{IKM,Z,MAR}). This situation means $d=0$, and therefore $n=D$. So, it is more useful to use the formula (\ref{U1odd}) of $U_{1}$ for $n$ odd, 
\be
U_{1} \left(\sigma; d=0 \right) =  - \frac{h}{D (4\pi)^{D/2} } \Gamma \left( 1 - \frac{D}{2} \right) \sigma^D .
\lb{U1d0}
\ee
Notice that $U_{1}$ diverges  for $D$ even \footnote{It is possible to see that the employment of the minimal subtraction scheme in Eq. (\ref{U1d0}) generates the same expression for $U_1$ if we have used Eq. (\ref{U1even}) with $d=0$.}. 
In particular, for $D=2$, four-fermion theories are renormalizable, and the renormalization is implemented by imposing the following renormalization condition, 
\be 
\left.\frac{\partial ^2 }{\partial  \sigma^2} U_{eff}\left(\sigma; d=0 \right) \right|_{\sigma = \mu} = \frac{1}{G_R}, 
\lb{GR}
\ee
which, with the help of Eqs. (\ref{effpot}) and (\ref{GR}), yields 
\be 
\frac{1}{G_R} = \frac{1}{G} -  \frac{h(D-1)}{ (4\pi)^{D/2} } \Gamma \left( 1 - \frac{D}{2} \right) \mu^{D-2} 
\lb{GR1}
\ee
Thus, we can write the renormalized effective potential, valid for $ 2 \leq D < 4$, as 
\be
\frac{1}{N} U_{eff} \left(\sigma; d=0 \right) =  \frac{\sigma ^2}{2G_R}+ \frac{h(D-1)}{ (4\pi)^{D/2} }  \Gamma \left( 1 - \frac{D}{2} \right) \mu^{D-2} \sigma ^2 - \frac{h}{D (4\pi)^{D/2} } \Gamma \left( 1 - \frac{D}{2} \right) \sigma^D .
\lb{effren}
\ee
Hence, the non-trivial solution of the gap equation is given by 
\be 
\frac{1}{G_1} = \frac{1}{G_0} +  \frac{h}{ (4\pi)^{D/2} } \Gamma \left( 1 - \frac{D}{2} \right) \left(\frac{m}{\mu}\right)^{D-2}, 
\lb{gap1}
\ee
where we have defined the dimensionless coupling constant $G_1 = \mu^{D-2} G_R$, and 
\be 
G_0 = \frac{(4\pi)^{D/2}}{h (1-D) \Gamma \left( 1 - \frac{D}{2} \right) }.
\lb{G0}
\ee
From Eq. (\ref{gap1}), we can infer that for $G_1 < G_0$ the allowed value of the mass $m$ is trivial, while for $G_1 > G_0$ the mass acquires a nonzero positive value. In this context, the constant $G_0$ acts as a critical value which $G_1$ must exceed to have dynamically generated fermion mass.

\subsection{The presence of boundaries}

In the case of presence of boundaries, we can use the same systematics as above, and therefore determine the finite-size contributions to the expression for $G_1$, with the only difference that we must analyze the gap equation in Eq. (\ref{gap1}) with its last term in right hand side in a different way,
\be 
\frac{1}{G_1} = \frac{1}{G_0} -  \frac{1}{ \bar{m} \mu ^{D-2}} \left.\frac{\partial  }{\partial \sigma}U_{1} \left(\sigma ;\left\{a_i\right\}, \left\{c_i\right\} \right)\right|_{\sigma = \bar{m}}
\lb{gap2}
\ee
where $\bar{m}$ is the fermion mass with the system in the presence of boundaries.

We can calculate the derivative of $U_{1}$ with respect to $\sigma$ directly from Eq. (\ref{U1}), by acting it on the zeta-functions. Taking into account the different possibilities for $U_{1}$, then the gap equation is written in the different situations, 
\ben 
\frac{1}{G_1} & = & \frac{1}{G_0} - \frac{h\mu ^{2-D}}{ V_d (4\pi)^{n/2} } \frac{(-1)^{\frac{n}{2}}}{\left(\frac{n}{2}-1\right)!} \left\{ Y_{d}^{\bar{m}^2 \;'}  \left( -\frac{n}{2}+1;\left\{a_i\right\}, \left\{c_i\right\} \right) \right. \nonumber \\
&+ & \left. Y_{d}^{\bar{m} ^2}\left( -\frac{n}{2}+1 ;\left\{a_i\right\}, \left\{c_i\right\} \right)\left[ \ln{\mu ^2} - \gamma - \psi \left(\frac{n}{2} \right) \right] \right\}, \;\;
\rm{ for \;\textit{n} \; even};
\lb{gap4}
\een
while
\be 
\frac{1}{G_1} = \frac{1}{G_0} +  \frac{h\mu ^{2-D}}{ V_d(4\pi)^{n/2} } \Gamma \left( 1 - \frac{n}{2} \right) Y_{d}^{\bar{m}^2 }\left( -\frac{n}{2}+1; \left\{a_i\right\}, \left\{c_i\right\} \right), \;\; \rm{ for \;\textit{n} \; odd};
\lb{gap3}
\ee
(remembering that Eq. (\ref{gap3}) is valid for $d$ even or in
the limit $\sigma ^2 \rightarrow 0$) and finally we have obtained,
\be 
\frac{1}{G_1} = \frac{1}{G_0} -  \frac{h\mu ^{2-D}}{ V_d } \; FP \; Y_{d}^{\bar{m}^2 }\left( 1; \left\{a_i\right\}, \left\{c_i\right\} \right), \;\; \rm{ for \;\textit{n}=0},
\lb{gap5}
\ee
in which we have considered only the relevant terms for further calculations; $FP$ means the finite part of $Y_{d}^{\bar{m}^2 }$. 

Now, we are able to determine the critical value of the coupling
constant $G_1$ with the corrections due to the presence of boundaries,
just by taking the fermion mass approaching to zero in the gap
equation. In this context, we can use the expansion in
Eq. (\ref{epstein3}) for $\bar{m} \rightarrow 0$; taking only the
first term of this expansion, then $Y_{d}^{\bar{m}^2 }$ reduces to the
homogeneous generalized Epstein zeta-function $Y_{d}$. However, as it
was remarked in the previous section , the pole structure of  $Y_{d}$
is simpler than that of  $Y_{d}^{\bar{m}^2 }$; therefore
Eq. (\ref{gap3}) becomes valid for general $d$.

We can construct analytical continuations for $Y_{d}$ to write it
in terms of Bessel and Riemann zeta functions, by considering a generalization of  recurrence formulas in \cite{EE},
\ben
Y_{d} \left( \nu; \left\{a_i\right\}, \left\{c_i\right\} \right)  =  \frac{\Gamma\left( \nu - \frac{1}{2}
 \right)}{ \Gamma(\nu)}  \sqrt{\frac{\pi}{a_d}} Y_{d-1}\left(\nu - \frac{1}{2};\left\{a_{j \neq d}\right\}, \left\{c_{j \neq d}\right\} \right) 
 +  \frac{4 \pi^s }{\Gamma(\nu)} W_d \left( \nu - \frac{1}{2}; \left\{a_i\right\}, \left\{c_i\right\} \right),
\label{epstein4}
\een
where the set $\left\{a_{j \neq d}\right\}$ means that the parameter $a_d$ is excluded from it, and 
\begin{equation}
W_d \left( \eta ; \left\{a_i\right\}, \left\{c_i\right\} \right) =  \frac{1}{\sqrt{a_d} }
\sum_{\left\{n_{j \neq d}\right\} \in \rm{\textbf{Z}}_0} \sum_{n_d=1}^{\infty}\cos{\left(2\pi n_d c_d \right)} \left( \frac{ n_d}{\sqrt{a_d} X_{d-1}}  \right)^{\eta}
K_{\eta}\left( \frac{2\pi n_d}{\sqrt{a_d} } X_{d-1} \right) \; , \label{Wd}
\end{equation}
with $X_{d-1}=\sqrt{\sum_{k=1}^{d-1}a_{k}\left(n_{k} + c_k \right)^{2}} $ and $K_{\nu} (z)$ being the modified Bessel function of second kind. 

Notice from Eq. (\ref{epstein4}) that whatever the sum one chooses to
perform firstly, when $c_i = c_j $ the $L_{i} \leftrightarrow L_{j}$
symmetry is lost \cite{Kirsten}; in order to preserve this symmetry, we adopt here a symmetrized summation generalizing the prescription explicit in
\cite{Abreu} for $c_i=0$. 

In the next section we will consider particular cases.

\section{Phase boundary}

In this section we will analyze the expressions (\ref{gap3})-(\ref{gap5}) for $\bar{m} \rightarrow 0$, which defines a critical surface $L_i$-dependent for the phase diagram of fermion mass. In the first three subsections, we consider the situations where all three dimensions are spatial coordinates. In the last subsection, we analyze the system at finite temperature, associating one of compactified dimension to the range of the inverse of temperature. 

\subsection{Case $d=1$, $n$ even }

We now consider the simplest case of the
compactification, namely when only one coordinate is compactified. This situation was in part studied in Ref. \cite{IKM} considering $L_1 \equiv \beta$; for completeness we will describe it here. 

For $d=1$, Eq. (\ref{eps}) is written as
\be
Y_{1} \left( \nu; a_1 \equiv a , c_1 \equiv c \right) = a^{-\nu} \left[ \zeta (2\nu, c ) + \zeta (2\nu, 1-c ) \right], 
\lb{Y1}
\ee
where $ \zeta (s, c ) = \sum _{k=0}^{\infty} \left[ k + c \right] ^{-s}$ is the Hurwitz zeta-function. The use of this last equation in (\ref{gap4}) for $\bar{m} \cong 0$ yields
\ben 
\frac{1}{G_1} & = & \frac{1}{G_0} - \frac{h\mu ^{2-D}}{ L (4\pi)^{\frac{D-1}{2} }} \frac{(-1)^{\frac{D-1}{2}}}{\left(\frac{D-3}{2}\right)!} \left\{ a^{\frac{D-3}{2}}  \left[ \zeta '(D-3, c ) + \zeta '(D-3, 1-c ) \right]  \right. \nonumber \\
&+ & \left. a^{\frac{D-3}{2}}\left[  \ln{\mu ^2} -\psi \left(\frac{D-1}{2} \right)- \gamma   - \ln{a} \right]  \left[ \zeta (D-3, c ) + \zeta (D-3, 1-c ) \right]   \right\}.
\lb{Gc1}
\een
Employing the expansion formula for the Hurwitz zeta-function, 
\be
\zeta (\nu, c ) \cong \frac{1}{2} - c + \nu \left[ \ln{\Gamma (c)} - \frac{1}{2} \ln{2\pi}\right], 
\lb{hurwitz-exp}
\ee
Eq. (\ref{Gc1}) becomes, for $D=3$,
\be 
\frac{1}{G_1}  =  \frac{1}{G_0} - \frac{1}{\pi L \mu } \ln{\left(\frac{2 \pi}{\Gamma (c)\Gamma (1-c)}\right)}. 
\lb{Gc2}
\ee

Notice that for $c=\frac{1}{2}$, Eq. (\ref{Gc2}) reduces to 
\be 
\frac{1}{G_1}  =  \frac{1}{G_0} - \frac{A_1}{ L \mu } , 
\lb{Gc22}
\ee
where $A_1 = \frac{ \ln2}{\pi} = 0.221 $. 

Noting that using (\ref{gap1}) in (\ref{Gc22}), we obtain the formula $L m = 2 \ln{2}$. It can be seen that the replacement of the length $L $ by the inverse of temperature $\beta$ gives the equation $T_c = \frac{m}{ 2  \ln{2}}$, which can be identified as the critical temperature of a second order phase transition, as has been remarked in Ref. \cite{Kim}.

Furthermore, yet with $L \equiv \beta$ and $c=\frac{1}{2} - \frac{i\beta \mu_0}{2 \pi}$, where $\mu_0$ is the chemical potential,  Eq. (\ref{Gc2}) becomes 
\be 
\frac{1}{G_1}  =  \frac{1}{G_0} - \frac{1}{\pi L \mu } \ln{\left[2\cosh{\left(\frac{\beta \mu_0}{2}\right)}\right]} . 
\lb{Gc3}
\ee
The use of Eq. (\ref{gap1}) in (\ref{Gc3}) generates the following formula for the critical temperature and the chemical potential, $\frac{\beta m}{2} = \ln{\left[2 \cosh{\frac{\beta \mu_0}{2}}\right]}$, which agrees with that obtained in Ref. \cite{IKM}.

\subsection{Case $d=2$, $n$ odd}

We now focus our interest in the case of two compactified coordinates. Taking $d=2$ in Eq. (\ref{gap3}) with $\bar{m} \cong 0$ generates 
\be
\frac{1}{G_1}  =  \frac{1}{G_0} + \frac{h\mu ^{2-D}}{ V_2 (4\pi)^{\frac{D-2}{2} }}
\Gamma \left(- \frac{D-4}{2}\right) Y_{2} \left( - \frac{D-4}{2} ; a_1, a_2, c_1, c_2 \right).
\lb{Gc4}
\ee
The analytical continuation of this equation can be performed by
using (\ref{epstein4}) in its symmetrized version,  (\ref{Y1}) and
(\ref{hurwitz-exp}), allowing us to rewrite Eq. (\ref{Gc4}) for $D=3$ as  
\ben
\frac{1}{G_1}  =  \frac{1}{G_0} - \frac{1 }{ \mu V_2 \pi }
\left\{ \frac{1}{2} \sum _{i=1}^{2}L_{ i} 
\ln{\left(\frac{2 \pi}{\Gamma \left(c_{j }\right)\Gamma \left(1-c_{j }\right)}\right)} 
 -  \sum _{ i=1  } ^{2} 
L_{ i} \sum_{n_i =1}^{\infty} \;\;\sum_{n_{j } =- \infty  }^{\infty} \cos{\left(2 \pi n_i c_i \right)} 
K_{0} \left( 2 \pi \frac{L_i}{L_{j}} n_i |n_{j} + c_{j} |\right) \right\} ,
\lb{Gc6}
\een
with $j \neq i$.

Considering $c_i = \frac{1}{2}$, we obtain ($j \neq i$)
\ben
\frac{1}{G_1}  =  \frac{1}{G_0} - \frac{1 }{  \pi }
\left\{ \left( \frac{1}{2L_1 \mu} + \frac{1}{2L_2 \mu}\right) \ln{2} 
 -   \sum _{i=1} ^{2} \frac{1}{L_{ j} \mu} \sum_{n_i =1}^{\infty} \sum_{n_j =- \infty}^{\infty} (-1)^{n_i} 
K_{0} \left( 2 \pi \frac{L_i}{L_{j}} n_i \left|n_{j} + \frac{1}{2} \right|\right) \right\}.
\lb{Gc7}
\een
In particular, for $L_1 = L_2 \equiv L$, we have 
\be 
\frac{1}{G_1} = \frac{1}{G_0} - \frac{A_2 }{ L \mu}, 
\lb{Gc8}
\ee
where 
\be
A_2=\frac{\ln2}{\pi} - \frac{2}{\pi}  \sum_{n_i =1}^{\infty} \sum_{n_2 =- \infty}^{\infty} (-1)^{n_1} 
K_{0} \left( 2 \pi  n_1 \left|n_{2} + \frac{1}{2} \right|\right) \cong 0.257 \;.
\lb{Gc88}
\ee

\subsection{Case $d=3$, $n=0$ }

We now pay our attention to the case where three dimensions are compactified. Taking $d=3$ in Eq. (\ref{gap5}), we obtain for $\bar{m} \cong 0$,
\be 
\frac{1}{G_1} = \frac{1}{G_0} -  \frac{2}{ \mu V_3 } \; FP \; Y_{3}\left( 1; a_1, a_2, a_3, c_1, c_2, c_3 \right) .
\lb{Gc9}
\ee

The analytical structure of the function
$Y_{3}\left( 1;  a_1, a_2, a_3, c_1, c_2, c_3 \right)$
can be obtained from the general symmetrized recurrence
relation given by Eq. (\ref{epstein4}); explicitly, one has
\begin{eqnarray}
Y_{3}\left( 1; a_1, a_2, a_3, c_1, c_2, c_3 \right)  = 
 \frac{\pi }{3} \sum_{i,j,k=1}^{3}\frac{(1+\varepsilon_{ijk})}{2}
\frac{1}{\sqrt{a_{i}}} Y_2 \left(\frac{1}{2};a_{j},a_{k},c_{j},c_{k}\right)
 + \frac{4\pi}{3} W_{3}\left(\frac{1}{2};a_{1},a_{2},a_{3}, c_1, c_2, c_3\right),
\label{E3}
\end{eqnarray}
where $\varepsilon_{ijk}$ is the totally antisymmetric symbol. With the same expression for $Y_2$ as used in Eq. (\ref{Gc4}), we can rewrite (\ref{Gc9}) as
\ben
\frac{1}{G_1}  &=&  \frac{1}{G_0} - \frac{1 }{3 \pi V_3  \mu}
\left\{  \sum _{i,j,k=1}^{3} \frac{(1+\varepsilon_{ijk})}{2}L_{ i}\left[  \frac{L_{ j}}{2}  \ln{\left(\frac{2 \pi}{\Gamma \left(c_{k }\right)\Gamma \left(1-c_{k }\right)}\right)} \right. \right.\nonumber \\
& - &  \left.  L_{ j} \sum_{n_j =1}^{\infty} \;\;\sum_{n_{k } =- \infty  }^{\infty} \cos{\left(2 \pi n_j c_j \right)} 
K_{0} \left( 2 \pi L_j n_j T_{k} \right)  + \left(L_j \leftrightarrow  L_k \right) \right] \nonumber \\ 
& - &  \left.   \sum _{i,j,k=1}^{3} (1+\varepsilon_{ijk})\;L_{ i}
\sum_{n_i =1}^{\infty} \cos{\left(2 \pi n_i c_i \right)} 
\sum_{n_{j }, n_{k } =- \infty  }^{\infty}
\left( \frac{ n_i L_i}{ T_{\,j \;k} } \right)^{\frac{1}{2}}
K_{\frac{1}{2}}\left( 2\pi n_i L_i T_{\,j\,k} \right) \right\}  , 
\lb{Gc10}
\een
where $T_{ij...}= \sqrt{\frac{(n_i+c_i)^2}{L_i^2}+\frac{(n_j+c_j)^2}{L_j^2} +...}$.

If we restrict ourselves to the situation where $c_i = \frac{1}{2}$, then we obtain the following equation that describes the critical coupling, 
\ben
\frac{1}{G_1} & =&  \frac{1}{G_0} - \frac{1 }{3 \pi V_3  \mu}
\left\{ \sum _{i,j,k=1}^{3} \frac{(1+\varepsilon_{ijk})}{2} L_{ i} \left[ \frac{L_{ j}}{2}\ln{2}  \right. \right.\nonumber \\
& - &  \left.  L_{ j}\sum_{n_j =1}^{\infty} \;\;\sum_{n_{k } =- \infty  }^{\infty} (-1)^{n_j}
K_{0} \left( 2 \pi L_j n_j \tilde{T}_{k} \right) +  \left(L_j \leftrightarrow  L_k \right)  \right] \nonumber \\ 
& - &  \left.   \sum _{i,j,k=1}^{3} (1+\varepsilon_{ijk})\;L_{ i}
\sum_{n_i =1}^{\infty} (-1)^{n_i} 
\sum_{n_{j }, n_{k } =- \infty  }^{\infty}
\left( \frac{ n_i L_i}{ \tilde{T}_{\,j\,k} }  \right)^{\frac{1}{2}}
K_{\frac{1}{2}}\left( 2\pi n_i L_i \tilde{T}_{\,j\,k} \right) \right\}  ,  
\lb{Gc101}
\een
where $\tilde{T} _{ij...} = \left.T _{ij...}\right|_{ c_i,c_j,... = \frac{1}{2} } $.

In particular, for $L_1 = L_2 = L_3 \equiv L$, we get 
\be 
\frac{1}{G_1} = \frac{1}{G_0} - \frac{A_3 }{ L \mu}, 
\lb{Gc11}
\ee
where  
\ben
A_3 &= & \frac{\ln2}{\pi} - \frac{2}{\pi}  \sum_{n_i =1}^{\infty} \sum_{n_2 =- \infty}^{\infty} (-1)^{n_1} 
K_{0} \left( 2 \pi  n_1 \left|n_{2} + \frac{1}{2} \right|\right)   \nonumber \\
& - &  \frac{2}{\pi}\sum_{n_1 =1}^{\infty} (-1)^{n_1} 
\sum_{n_{2 }, n_{3 } =- \infty  }^{\infty}
\left( \frac{ n_1 }{ \sqrt{\sum_{i=2}^{3}\left(n_i+\frac{1}{2}\right)^2}}  \right)^{\frac{1}{2}}
K_{\frac{1}{2}}\left( 2\pi n_1 \sqrt{\sum_{i=2}^{3}\left(n_i+\frac{1}{2}\right)^2} \right)
 \cong  0.278 .
\lb{Gc12}
\een

\begin{figure}[th]
\includegraphics[{height=5.0cm,width=7.5cm}]{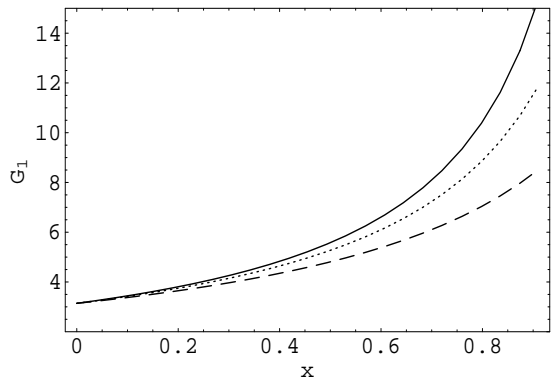}
\caption{ The critical coupling constant $G_1$ as a function of
  $\texttt{x}= (L \mu )^{-1}$  for the three situations: dashed,
  dotted and solid lines  represent Eqs. (\ref{Gc22}), (\ref{Gc88})
  and (\ref{Gc11}), which  are the cases of $d=1,2,3$ compactified
  dimensions, respectively, $L$ being the length in each case. For
  each situation the  non-trivial mass phase corresponds to the region
  above the corresponding line. }
\label{FIG1}
\end{figure}

In Fig. \ref{FIG1} is plotted the critical coupling constant $G_1$ as a function of $\texttt{x} = (L \mu )^{-1}$ representing Eqs. (\ref{Gc22}), (\ref{Gc88}) and (\ref{Gc11}). These contexts correspond, respectively, to the system in the following scenarios: confined between two parallel planes a distance $L$
apart; confined to an infinitely long cylinder
having a square transversal section of area $L^2$; and to a cubic
box of volume $L^3$. We can see that as $L$ decreases, $G_1$ increases monotonically, meaning that to keep the non-trivial mass phase with compactified coordinates, it is necessary a stronger interaction than that in non-compactified space. 

In addition, from Fig. \ref{FIG1} it is easy to see that there is a
critical value $L_c$ below which there exists no more non-vanishing
fermion mass, since the coupling constant tends to infinity at this
point. We have the following values for $\texttt{x} _c \equiv (L_c \mu )^{-1}= 1.44$ for $d=1$, 1.24 for $d=2$ and 1.14 for $d=3$. Thus, the cubic box gives the greatest value for $L_c$, which means that the increasing of the number of compactified dimensions forces a stronger interaction to stay in the region of non-trivial mass.

\subsection{System at finite temperature}

Now we turn our attention to the system with compactified dimensions and at finite-temperature. First, let us look at the situation in Eq. (\ref{Gc7}), which has $d=2$. Taking $L_1 \equiv L$ and $L_2 \equiv \beta$, we obtain the expression for $G_1$ which is dependent of the temperature effects and of the size of the compactified dimension. However, we also can use Eq. (\ref{gap1}), obtaining 
\ben
&-&\frac{m}{\mu} + 
\left( \frac{1}{L \mu} + \frac{1}{\beta \mu}\right) \ln{2} -  \frac{2}{L\mu}  \sum_{n_1 =1}^{\infty} \sum_{n_2 =- \infty}^{\infty} (-1)^{n_1} 
K_{0} \left( 2 \pi \frac{\beta}{L} n_1 \left|n_{2} + \frac{1}{2} \right|\right)
\nonumber \\
& - &   \frac{2}{\beta \mu} \sum_{n_1 =1}^{\infty} \sum_{n_2 =- \infty}^{\infty} (-1)^{n_1} 
K_{0} \left( 2 \pi \frac{L}{\beta} n_1 \left|n_{2} + \frac{1}{2} \right|\right) =0.
\lb{PD1}
\een 

\begin{figure}[th]
\includegraphics[{height=7.0cm,width=7.0cm}]{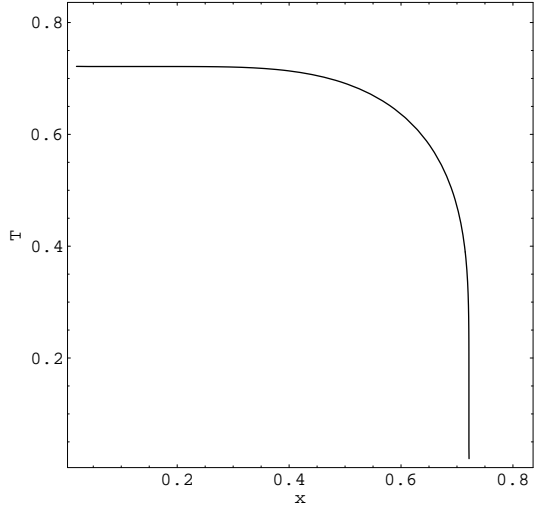}
\caption{ The phase diagram corresponding to Eq. (\ref{PD1}) (finite $T$ and one compactified spatial dimension) in the $\texttt{x}-\texttt{T}$ plane, where $\texttt{x}= (L m )^{-1}$ and $ \texttt{T} = (\beta m )^{-1}$. The non-zero mass phase corresponds to the region below the curve.}
\label{FIG2}
\end{figure}

The corresponding phase diagram in the $\texttt{x}-\texttt{T}$ plane of Eq. (\ref{PD1}) is
plotted in Fig. \ref{FIG2},  where $\texttt{x}= (L m )^{-1}$ and $
\texttt{T} = (\beta m )^{-1}$.  Notice that the critical line
separates the non-vanishing  fermion mass phase below the line and the
trivial  fermion mass phase above the line. As we can see,  the size decreasing reduces the critical temperature. Also, the diagram has a manifest $\texttt{x} \leftrightarrow \texttt{T} $ symmetry.

Furthermore, in the $L \rightarrow \infty$ limit, which in the diagram
means $\texttt{x} \rightarrow 0$, the variable $ \texttt{T}$ reaches
the usual critical value $\texttt{T}_c= 1/(2\ln{2})$. Analytically,
this results from Eq. (\ref{PD1}), in which only the first term with Bessel function survives, and the summation over $n_2$ is replaced by an integration, yielding 
\ben
\frac{2}{L\mu}  \sum_{n_1 =1}^{\infty} \sum_{n_2 =- \infty}^{\infty} (-1)^{n_1} 
K_{0} \left( 2 \pi \frac{\beta}{L} n_1 \left|n_{2} + \frac{1}{2}
  \right|\right) &  \stackrel{L \rightarrow \infty}{\longrightarrow}
& 
  \frac{4}{\mu} \sum_{n =1}^{\infty}(-1)^{n} \int _{0}^{\infty} \frac{d p}{2 \pi} K_0 \left( n \beta p \right)  \nonumber \\
& = & - \frac{1}{ \beta \mu } \lim_{s \rightarrow 1} \left(1 -2^{1-s}\right) \zeta \left( s \right)
 = - \frac{1}{ \beta \mu } \ln{2}. 
\lb{limit}
\een 
Hence, the replacement of Eq. (\ref{limit}) in (\ref{PD1}) in the
limit $L \rightarrow \infty$ reproduces the correct critical value $\texttt{T}_c= 1/(2\ln{2}) = 0.72$. Moreover, since the diagram has a $\texttt{x} \leftrightarrow \texttt{T} $ symmetry, the same argument is applied to $L $ in the limit $\beta \rightarrow \infty$.
\\

Now we analyze the case with three compactified dimensions, by using
Eq. (\ref{gap1}) in Eq. (\ref{Gc101}) and taking $L_1=L_2 \equiv L$
and $L_3 \equiv \beta$. Then, we obtain the following equation for critical line, 
\ben
&-&\frac{m}{\mu} +\frac{2}{3} 
\left( \frac{2}{L \mu} + \frac{1}{\beta \mu}\right) \ln{2} -  \frac{4}{3L\mu}  \sum_{n_1 =1}^{\infty} \sum_{n_2 =- \infty}^{\infty} (-1)^{n_1} 
K_{0} \left( 2 \pi \frac{\beta}{L} n_1 \left|n_{2} + \frac{1}{2} \right|\right)  \nonumber \\
& - &  \frac{4}{3\beta \mu} \sum_{n_1 =1}^{\infty} \sum_{n_2 =- \infty}^{\infty} (-1)^{n_1} 
K_{0} \left( 2 \pi \frac{L}{\beta} n_1 \left|n_{2} + \frac{1}{2} \right|\right) - \frac{4}{3L \mu} \sum_{n_1 =1}^{\infty} \sum_{n_2 =- \infty}^{\infty} (-1)^{n_1} 
K_{0} \left( 2 \pi  n_1 \left|n_{2} + \frac{1}{2} \right|\right) \nonumber \\ 
& - &  \frac{8}{3L \beta \mu} \sum_{n_1 =1}^{\infty} (-1)^{n_1} 
\sum_{n_{2 }, n_{3 } =- \infty  }^{\infty}
\left( \frac{ n_1 L}{ \tilde{T}_{L \beta}}  \right)^{\frac{1}{2}}
K_{\frac{1}{2}}\left( 2\pi n_1 L \tilde{T}_{L \beta} \right) \nonumber \\
& - &  \frac{4}{3L^2 \mu} \sum_{n_1 =1}^{\infty} (-1)^{n_1} 
\sum_{n_{2 }, n_{3 } =- \infty  }^{\infty}
\left( \frac{ n_1 \beta }{ \tilde{T}_{L L} } \right)^{\frac{1}{2}}
K_{\frac{1}{2}}\left( 2\pi n_1 \beta \tilde{T}_{L L} \right) =0 .
\lb{PD2}
\een

\begin{figure}[th]
\includegraphics[{height=7.0cm,width=7.0cm}]{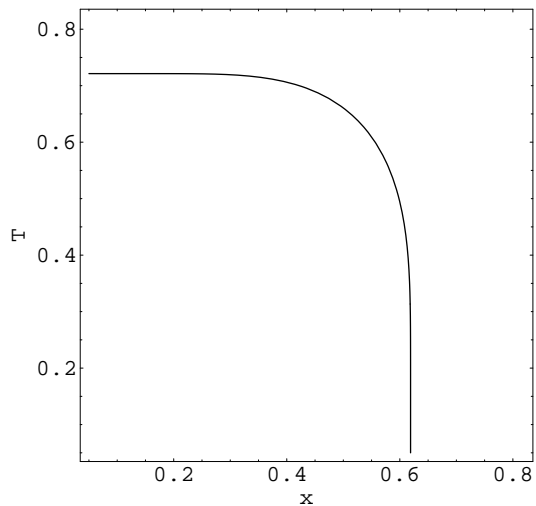}
\caption{ The phase diagram corresponding to
  Eq. (\ref{PD2}) (finite $T$ and two compactified spatial dimensions) in the $\texttt{x}-\texttt{T}$ plane; $\texttt{x}= (L m )^{-1}$ and $ \texttt{T} = (\beta m )^{-1}$. The non-zero mass phase corresponds to the region below the curve. }
\label{FIG3}
\end{figure}

The corresponding phase diagram of Eq. (\ref{PD2}) in the $\texttt{x}-\texttt{T}$ plane is plotted in Fig. \ref{FIG3}. As we can see, there is no $\texttt{x} \leftrightarrow \texttt{T} $ symmetry in the case of the critical line of Eq. (\ref{PD2}). Besides, the critical temperature has a faster decreasing with the size reduction than the case of Fig. \ref{FIG2}. 

It should be noted that, by employing the same systematics used to get
Eq. (\ref{limit}), the $\texttt{x} \rightarrow 0$ gives again
$\texttt{T}_c= 1/(2\ln{2}) = 0.72$. On the other hand, the
$\texttt{T} \rightarrow 0$ limit generates $\texttt{x}_c \cong
0.62$. 

\section{Concluding Remarks}

In this paper, we have discussed the finite-size effects on the phase structure of the Euclidean three-dimensional NJL model in the frame of zeta function and compactification methods. 

In the leading order of the $\frac{1}{N}$-expansion, we analytically derived the expressions for the renormalized effective potential and gap equation for different situations of compactified coordinates without any further approximation. 

Taking first the situation in which the dimensions are spatial and
with antiperiodic boundary conditions, we estimated the strength of
the critical coupling constant for the cases of one, two and three
compactified dimensions with the same length $L$ and compare them. We found
that increasing of the number of compactified dimensions
requires the increasing of the strength of the interaction to remain in the region of non-trivial mass. 

We also investigated the system at finite temperature, which formally
is done by associating the size of one of compactified dimensions to
$\beta = 1/T$. We have obtained the phase diagram in terms of
temperature and inverse of the size and noted that the phase transition
is easier (bigger values of $L$) for the situation with the largest number of compactified coordinates. 

It is worthy to mention that we have obtained general expressions for
the effective potential and gap equations (see Eqs. (\ref{U1odd})-(\ref{U1even}) and (\ref{gap3})-(\ref{gap5})), for different numbers of total and compactified dimensions and also for different
choices of the $c_i$-parameters that appear in
Eq. (\ref{prescription}). This approach in principle allows us to
study other cases, as for example the phase diagram at finite chemical potential as well as non-trivial topologies. 
\begin{acknowledgments}
This work was partially supported by Funda\c{c}\~{a}o de Amparo \`{a} Pesquisa do Estado de S\~{a}o Paulo (FAPESP) and Conselho 
Nacional de Desenvolvimento Cient\'{\i}fico e Tecnol\'{o}gico (CNPq).
The work of L. M. Abreu. was supported by FAPESP,  project 05/51099-0. 
\end{acknowledgments}

\end{document}